# Deadline-aware Scheduling for Maximizing Information Freshness in Industrial Cyber-Physical System

Devarpita Sinha, *Graduate Student Member, IEEE* and Rajarshi Roy, *Senior Member, IEEE*

*Abstract*—"Age of Information" is an interesting metric that captures the freshness of information in the underlying applications. It is a combination of both packets inter-arrival time and packet transmission delay. In recent times, advanced real-time systems rely on this metric for delivering status updates as timely as possible. This paper aims to accomplish optimal transmission scheduling policy to maintain the information freshness of real-time updates in the industrial cyber-physical systems. Here, the coexistence of both cyber and physical units and their individual requirements to provide the quality of service is one of the critical challenges to handle. A greedy scheduling policy called *Deadline-aware highest latency first* has been proposed for this purpose. This paper also gives the analytical proof of its optimality, and finally, the claim is validated by comparing the performance of our algorithm with other scheduling policies by extensive simulations.

*Index Terms*— Age of Information, Information Freshness, Industrial Cyber-Physical System, Greedy Scheduling, Utility of Information, Packet Deadline, Latency, Jitter.

## I. INTRODUCTION

Age of Information (AoI) [1] is a newly proposed cross-layer metric to measure the freshness of information from the receiver perspective. It grows linearly with the time elapsed since the generation of the latest packet received [1]. Unlike traditional packet delay, AoI ensures the regular delivery of time-varying updates with minimum end to end delay. This unique feature enables AoI to improve the quality of service (QoS) of the system from both communication and control points of view [2]. In recent times, time-critical information updates are gaining ardent importance with the growth of the advanced technologies in a diverse field of real-time applications such as traffic, stock market or weather forecasts, object tracking, internet of things (IoT) based remote health monitoring, smart manufacturing using cyber-physical system (CPS), device-to-device (D2D) communications for autonomous vehicles, power management in smart grids and many more.

### A. Background and Motivation

CPS is a closed-loop networked control system that integrates computation, communication, and control of physical processes over a common wireless network. Application of CPS in the industry (e.g., smart factory) consists of a remotely sited wireless sensor-actuator network (WSAN) connected to some centralized or distributed controller(s) through high-speed internet. Sensors are sensing the .physical states of the plant and transmitting those real-time updates to the controller. Controllers are responsible for decision making and they are remotely managing the underlying tasks by employing the suitable actuators in action. *Our work is motivated by the need of the CPS for exchanging the time-sensitive fresh information through the WSAN in order to monitor, estimate, and closed-loop control of the plant state, remotely.* Failing this may lead to some production loss or fatal accidents. But, constraints like limited bandwidth and power resources, interference in wireless channels, etc. making the scheduling decision more complicated and challenging.

### B. Related Work

There is a variety of literature available that deal with minimum AoI scheduling problems for information freshness [3-11]. Paper [3] addresses optimal scheduling strategy for a multi-source, common channel wireless network with the objective of minimizing the overall information age for delivering the information as timely as possible. Whereas, paper [4] finds a decision policy that determines both the sampling times and transmission order of the sources for minimizing the total average peak age (TaPA) and the total average age (TaA) in a multi-source, common channel system. Scheduling problems for minimizing the average and peak AoI in wireless networks under general interference constraints and throughput constraints are considered in [5] and [6], respectively. Paper [7] optimizes age-of-information without throughput loss for a multi-server queue. It proved that preemptive Last Generated First Served (LGFS) policy provides age optimality for both infinite and finite buffer queues. However, the optimality of newly proposed algorithms preemptive and non-preemptive Maximum Age Difference (MAD) has been investigated in [8] for multi-flow single server networks with flow diversity. Markov decision process (MDP) is a strong tool for discrete-time decision making depending on the state of the environment. Reference [9-10] took the help of MDPs for optimal sampling and scheduling of updates in order to minimize the age of information for maximizing the data freshness. Reference [9] formulates a Markov decision process (MDP) to find dynamic transmission scheduling schemes for a wireless broadcast network, with the purpose of minimizing the long-run average age. At first, it shows that an optimal scheduling algorithm for MDP is a simple deterministic, stationary switch-type algorithm. Next, this paper proposes a sequence of finite-state approximations equivalent to infinite-state MDPs and finds both optimal off-line and online scheduling algorithms for these finite-state MDPs depending on the knowledge of time-varying packet arrivals. Paper [10] studies scheduling problem to collect fresh data in time-varying networks with power-constrained users. This can be done by decoupling the multi-user scheduling problem into a single user constrained Markov decision process (CMDP) and finally, the optimum solution is found through the linear programming (LP) method. Paper [11-12] investigates the optimal sampling problem for age minimization with the help of constrained continuous-time uncountable state MDP and semi-MDP, respectively. On the contrary, references [13-14] formulates a decision problem to find an optimum transmission scheduling policy to minimize the AoI of the destinations in a broadcast network. Works, done in [15-17], investigate scheduling algorithms to maximize the information freshness in terms of AoI in CPS, but they do not consider deadlines for packet transmissions. In contrast, [18-19] find scheduling policies to maintain the freshness of real-time updates in CPS, but they do not consider AoI as their freshness indicator. Reference [18] deals with the temporal validity of real-time data of a multi-modal behavioral problem in dynamic cyber-physical systems (DCPS). This proves that a utilization-based scheduling selection (UBSS) strategy can significantly outperform a single fixed update algorithm in terms of maintaining a better balance between data freshness and system schedulability. Paper [19] proposes two dynamic priority-based scheduling policy called jitter-based earliest deadline first (JB-EDF) and enhanced JB-EDF (JB-EDF*) to maintain the data freshness by reducing the sensor update workload for small scale and large scale update transactions, respectively, in distributed CPSs.

### C. Contribution

In this paper, we consider an industrial cyber-physical system (ICPS), engaged in monitoring and controlling a set of stochastic processes in a plant with the help of multiple centrally operated sensor-actuator pairs. Due to bandwidth limitation, multiple sensors are trying to transmit through a common shared channel. Now, these real-time updates have their own deadlines depending on the time-varying nature of the underlying processes. However, the age of information content present in any update increases linearly with time, and as a result, its freshness drops. Here, both the packet loss as well as outdated information degrade the communication and control performances of the overall system. Keeping these challenges in mind, this paper:

- Relates AoI with the latency and hence, freshness of a sensor sample.
- Formulates a utility function considering the deadline and freshness of a sensor sample.
- Proposes an age-based, deadline-aware packet scheduling policy to maximize the expected utility of the system. This policy, in turn, jointly maximizes the QoS (in terms of AoI, latency, and jitter) for the real-time monitoring and control of the industrial wireless sensor-actuator network (IWSAN).
- Proves the optimality of the algorithm analytically and compares its results with other popular algorithms in this context.

These fundamental contributions make our work unique from the existing literature.

### D. Outline of the paper

The rest of the paper is organized as follows. In Section II, we describe the system model and its related mathematical expressions are formulated in Section III. Section IV presents our proposed scheduling algorithm and proves its optimality through mathematical analysis. In Section V, we present and discuss the results of our simulation study. And last but not least, Section VI concludes the importance of the paper. However, in this paper, the words like sensor sample, status update, and packet are used interchangeably. The list of useful notations is presented in Appendix B.

## II. SYSTEM MODEL

### A. Basic Modelling

The system model, in this paper, consists of a symmetric IWSAN with M sensor-actuator pairs. Sensors and actuators are geographically distributed but wirelessly connected in a closed-loop through a centralized processor or controller, as shown in Fig. 1. Index for sensors and their corresponding actuators are represented with notation $i$, for $i = 1,2 \dots M$.

Sensors are updating the controller from time to time about the status of some physical parameters of their interests inside a plant. The Controller then analyzes that information and sends instructions to the corresponding actuators with a goal to handle the necessary control actions in the best possible way. In this model, sensors are transmitting their packets through a TDMA shared (according to WirelessHART standard [20] for IWSAN), single hop, time-varying, unreliable channel. When this channel is transmitting reliably, it is considered to be in 'ON' state. This 'ON' probability of the channel is $p_i = p \in [0,1]$. In contrast, when the channel fails to transmit it is said to be 'OFF' with probability ($1-p_i$). However, this status update task is limited by a finite time horizon $T$, and within this limit, each of the individual TDMA slots from time $[t-1,t]$ is denoted by $t$, for $t = 1,2,...,T$.

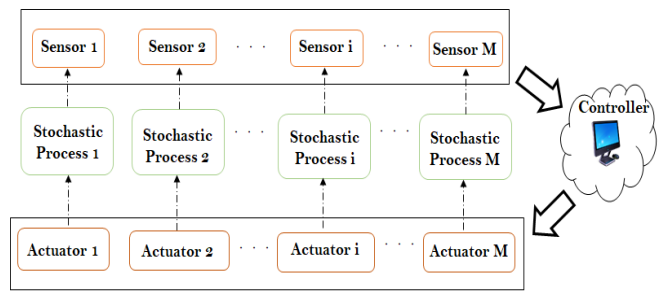

Fig. 1. Symmetric IWSAN with M sensors, M actuators and a centralized controller.

### B. Functionality

At the beginning of any slot $t$, sensors ready with their packets (we will call them 'active' sensors later in this paper) competes for sharing their status with the controller. The timing diagram of the controller is shown in Fig. 2. The controller schedules the transmission from one of the active sensors or decides to remain idle based on some designated scheduling policy π. After the sample from sensor $i$ reaches the controller, the controller generates appropriate control command by analyzing the information content present in that sample and reliably send them to the corresponding actuator $i$ by the end of the current slot. At the end of the successful processing of a sensor sample, the controller instantly and reliably sends feedback to the source sensor. So, starting from the sample collections by the sensors, scheduling decision making, analysis of the sample, control command generation, and up to the reception of feedback, the whole processing of a sensor sample takes one slot time in total.

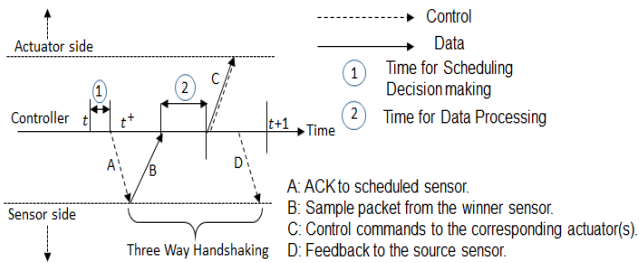

Fig. 2. Timing diagram of the controller.

### C. Age Evolution

Age of Information (AoI) of $k_i^{th}$ sample from sensor $i$ at the beginning of any time slot $t$ is indicated as $h_{t,i}(k_i)$ or simply $h_{t,i}$. Here, $k_i = 1,2,3,...$ is the real positive integer that denotes the index number of the incoming sample packet from sensor $i$. Starting from the generation of the latest sample received, the age of information increases linearly until the successful processing of the next sensor samples. This is shown in Fig 3. After the successful processing of a sensor sample at any slot $t$, in the beginning of the next slot $t+1$ the incoming packet index changes from $k_i$ to $k_i + 1$ and its AoI $h_{t+1,i}(k_i + 1)$ drops to a value, equals to the time passed after the last-time replacement of the latest successfully processed sample $k_i$, according to the definition of AoI [1]. This time is nothing but the processing time of $k_i$ and it is represented as of $P_i(k_i)$. In this model, $P_i(k_i)$ is always considered as one slot time (from Sec II. B). So, it can be said that after the successful processing of any sensor sample, the age of the next incoming sample from the same sensor drops to 1 at the beginning of the next slot, as shown in Fig. 3. On the contrary, unprocessed 'active' sensors, sense the present status of the system periodically in each slot by replacing the old samples and participate in scheduling. Their ages keep on increasing by one slot, each time they fail to transmit their samples. Therefore, in other words, **our definition of 'age of information' indicates the time passed after the collection of the last successfully processed sensor sample containing the information about the plant's state and its evolution with time can be expressed as,**

$$h_{t+1,i}(k_i) = h_{t,i}(k_i) + 1 : If\ i\ is\ not\ served\ at\ slot\ t$$

$$h_{t+1,i}(k_i + 1) = P_i(k_i) = 1 : If\ i\ is\ served\ at\ slot\ t \quad (1)$$

After receiving the control command, the actuator takes some time equivalent to next $c_i(k_i) \geq 0$ number of slots to finish the appropriate actuation task. During this period, no new actuation task is required to be invoked, and only after that period, the sensor resumes sensing. So, before age value $(c_i(k_i) + P_i(k_i)) = (c_i(k_i) + 1)$ is attained, sensor $i$ is said to be 'Inactive.' Inactive mode protects the resource-constrained network from unnecessary congestion and energy dissipation by avoiding redundant sensor updates. When the age of the sensor $i$ goes beyond the value $(c_i(k_i) + 1)$, this sensor becomes 'Active' and participates in scheduling until its newly sensed sample $k_i + 1$ is being processed successfully. But, if the $k_i^{th}$ sample of the active sensor $i$ fails to get service within its stipulated deadline $D_i(k_i)$, which is also known as the absolute deadline of a sample packet, then the packet from sensor $i$ is dropped and the $i^{th}$ flow-line (sensor-actuator pair) goes 'Out of service' permanently up to the end of $T$. This is, of course, not a desirable situation to occur. However, the underlying incomplete task of the broken flow-line is controlled (and eventually shut down, if applicable) by some default or preset control commands.

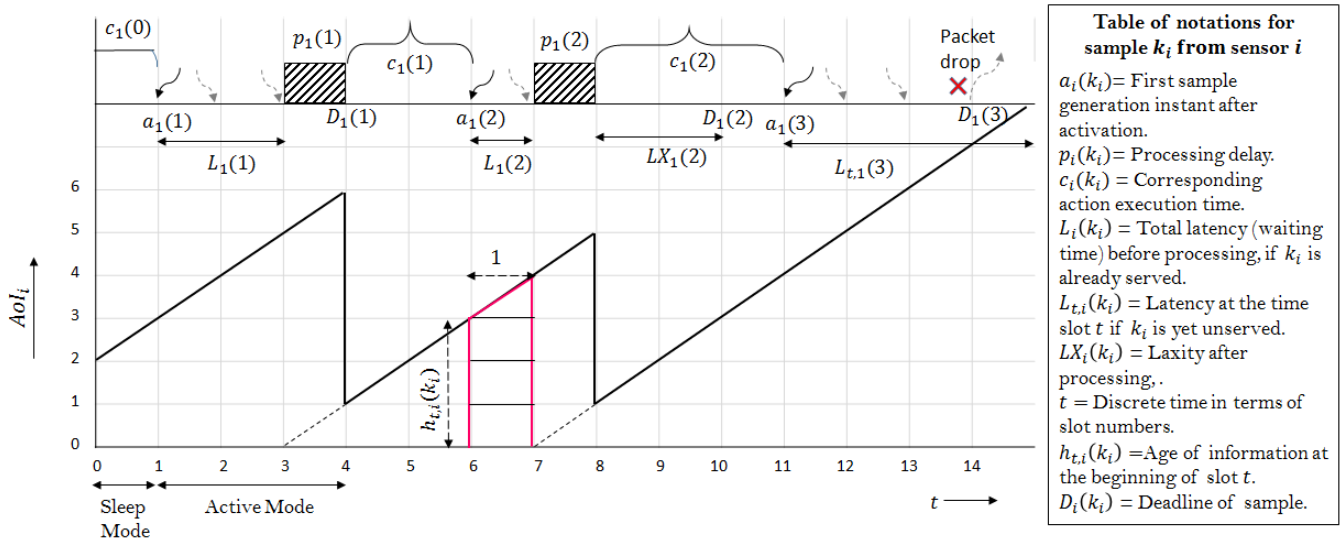

Fig. 3. Age evolution for sensor $i$.

So, at any slot $t$, the age of the sample $k_i$ from sensor $i$ can be represented as follows,

$$h_{t,i}(k_i) = \underbrace{P_i(k_i - 1) + c_{t,i}(k_i - 1)}_{Useful\ Age} + \underbrace{L_{t,i}(k_i)}_{Stale\ age}. \quad (2)$$

Here, $P_i(k_i - 1)$ is the processing time required to analyze the previous sample $k_i - 1$ from the same sensor $i$ and to generate the control command based on that sample; $c_{t,i}(k_i - 1)$ is the time required so far to perform the control action based on that control command based on the analysis of the previous sample from the same sensor. These two components of age are useful and unavoidable. Rest, the latency $L_{t,i}(k_i)$ at the present slot $t$, unnecessarily degrades the QoS of the sensor-actuator network by enhancing the end to end communication delay.

When any sensor $i$ with packet index $k_i$ is inactive at $t$, it means that the previous sensor sample $k_i - 1$ is successfully processed, but its corresponding actuation task is not completed yet. Then, as shown in Fig. 3, the sensor is yet to sense the next sample $k_i$ and, as a result, the latency of that sample $L_{t,i}(k_i) = 0$, for obvious reason. On the other hand, the active state of sensor $i$ signifies that the actuation task corresponding to its previously processed sample $k_i - 1$ is complete and the present sample $k_i$ is waiting for getting service by the processor. So, for the active sensor, $c_{t,i}(k_i - 1) = c_i(k_i - 1)$ and $L_{t,i}(k_i) \geq 0$.

Now, from the above discussion we can say that, at any time $t$, sensors lie in one of the three following modes,

$$Inactive: \quad h_{t,i}(k_i) < c_i(k_i - 1) + 1 \quad (3)$$

$$Active: \quad c_i(k_i - 1) + 1 \leq h_{t,i}(k_i) < D_i(k_i) \quad (4)$$

$$Out\text{-}of\text{-}Service: \quad c_i(k_i - 1) + 1 < D_i(k_i) \leq h_{t,i}(k_i) \quad (5)$$

## III. MATHEMATICAL FORMULATIONS

From Fig 3, the total age $AoI_i$ of an individual sensor $i$ is obtained from the area covered by the age curve for sensor $i$. It is expressed as follows,

$$AoI_i = \sum_{t=1}^{T} AoI_{t,i} = \sum_{t=1}^{T} \left\{ \frac{1}{2} + h_{t,i}(k_i) \right\}. \quad (6)$$

By ignoring the constant term in (6), the average age becomes,

$$\overline{AoI} \approx \frac{1}{TM} \sum_{t=1}^{T} \sum_{i=1}^{M} \{ h_{t,i}(k_i) \}$$

$$\approx \frac{1}{TM} \sum_{t=1}^{T} \sum_{i=1}^{M} P_i(k_i - 1) + c_{t,i}(k_i - 1) + L_{t,i}(k_i). \quad (7)$$

In (7), the two terms $P_i(k_i - 1)$ and $c_{t,i}(k_i - 1)$ are system defined and can not be controlled from outside. Hence, neglecting these two terms does not affect the original optimization problem for obtaining the information freshness. Next, considering only the variable term under control in (7), the average latency (AL) present in $\overline{AoI}$ is obtained as,

$$AL = \frac{1}{TM} \sum_{t=1}^{T} \sum_{i=1}^{M} \{ L_{t,i}(k_i) \}. \quad (8)$$

Communication delay $\Delta_i(k_i)$ of sample $k_i$ is dependent on the latency and measured as, $\Delta_i(k_i) = [L_i(k_i) + P_i(k_i)]$. Here, $L_i(k_i)$ is the total latency incurred, before serving any sample $k_i$. If the packet is yet unserved at $t$, the delay is said to be at least $\Delta_i(k_i) = [L_{t,i}(k_i) + P_i(k_i)]$. The standard deviation of the communication delay is known as RMS jitter.

The jitter is a metric associated with the actuation tasks, depending on which the QoS of the system is being measured. RMS jitter in the actuation tasks is obtained as,

$$\overline{JT} = \sqrt{\frac{\sum_{i=1}^{M} \sum_{k'=1}^{k_i} \{\Delta_i(k') - \overline{\Delta}\}^2}{\sum_{i=1}^{M} k_i}}$$

$$\text{where, the average delay } \overline{\Delta} = \frac{\sum_{i=1}^{M} \sum_{k'=1}^{k_i} \Delta_i(k')}{\sum_{i=1}^{M} k_i}. \quad (9)$$

From the existing literature, although it is well known that the smaller the average age, the higher the information freshness obtained. But by analyzing (2) and (6-7) referring to the Fig. 3, it can be said that the latency is an intensive part of the 'age,' and it's the only variable part in the 'age' term that can be controlled from outside. Thus, minimizing the average latency (AL) over the average age ($\overline{AoI}$) is crucial for this system model in order to maximize the information freshness of the sample packets. Moreover, a packet drop leads to the breakdown of the corresponding flow-line. **So, in order to maximize the information freshness in IWSAN, as mentioned in this work, two challenges are involved, as follows:**

1. Minimization of the average latency of sensor samples.
2. Minimum packet drops.

*By maximizing the new metric average (expected) utility of the information content present in the active sensor samples of the system, both the abovementioned challenges can be overcome and our objective of attaining maximum information freshness can be fulfilled.*

The utility of the information content in a sample $k_i$ from an active sensor $i$ is expressed as,

$$U_{t,i}(k_i) = f\left(F_{t,i}(k_i), X_{t,i}(k_i)\right)$$

$$where, X_{t,i}(k_i) = \mathcal{F}\left(LX_{t,i}(k_i)\right). \quad (10)$$

$F_{t,i}(k_i)$ is the freshness of information. It monotonically decreases with increasing latency $L_{t,i}(k_i)$ of the sensor sample $k_i$ that contains the information.

Latency is the number of time slots spent in the system by an active sensor sample before it is getting service successfully. From Fig. 3, the latency of any $k_i^{th}$ sample at slot $t > a_i(k_i)$ will be, $L_{t,i}(k_i) = (t - 1 - a_i(k_i))$ slots time, if this sample is not processed yet at slot $t$. On the other hand, if the sample $k_i$ is getting service at any slot $t$ and $a_i(k_i) < t \leq D_i(k_i)$, then the total latency incurred by the sample before processing is $L_i(k_i) = L_{t,i}(k_i) = (t - 1 - a_i(k_i))$.

When latency is zero, the freshness of the information is maximum ($F_{t,i}(k_i) = 1$). In this paper, $F_{t,i}(k_i)$ is considered as, $\frac{1}{L_{t,i}(k_i) + P_i(k_i)} = \frac{1}{Stale\ age + processing\ time}$. As $P_i(k_i) = 1$ in this system mode, therefore, $F_{t,i}(k_i) = \frac{1}{L_{t,i}(k_i) + 1}$.

Laxity $LX_{t,i}(k_i)$ of an active sensor sample $k_i$ is defined as the number of time slots left up to its deadline after the completion of the processing of that sample starting from the current instant. From Fig 3, laxity of $k_i^{th}$ active sample at slot $t$ is, $LX_{t,i}(k_i) = (D_i(k_i) - t)$. If $d_i(k_i)$ be the relative deadline of the $k_i^{th}$ sample and measured as $d_i(k_i) = D_i(k_i) - a(k_i)$, then laxity $LX_{t,i}(k_i)$ can be expressed as $LX_{t,i}(k_i) = (d_i(k_i) - L_{t,i}(k_i) - P_i(k_i))$ slots time.

When a sample is not scheduled for transmission even after having $LX_{t,i}(k_i) = 0$, then it is going to be dropped at the end of the current slot. This type of sample with zero laxity is called '*Critical Sample*.' If more than one critical samples are present at a particular slot $t$, only the critical sample with the highest utility value is considered to have a hard deadline. Other critical samples are going to be graced by delaying their deadlines by one more slot. As a result, their laxity becomes 1, and they turn out to be non-critical samples in the current slot. The same phenomena will take place to all the critical samples when the channel is *OFF*. This is called the **'Conflict Avoidance'** mechanism. So, from this, we can conclude that at most one critical sample can be present in the network at any time slot. However, an active sensor having the critical sample is called '*Critical Sensor*' and other active sensors with non-critical samples are called '*Non-critical Sensors*.'

Whenever a critical sample is dropped, its utility becomes 0 forever. However, any positive laxity of a sample does not affect its utility. So, the influence of laxity on utilization is captured in a variable denoted as $X_{t,i}(k_i)$ and it can be expressed as follows,

$$X_{t,i}(k_i) = \begin{cases} 0 & for\ LX_{t,i}(k_i) < 0 \\ 1 & for\ LX_{t,i}(k_i) \geq 0. \end{cases} \quad (11)$$

$U_{t,i}(k_i)$ is directly proportional to any real powers of $F_{t,i}(k_i)$ and $X_{t,i}(k_i)$. So, the utility can be expressed as,

$$U_{t,i}(k_i) = k_F k_X F_{t,i}^\beta(k_i) X_{t,i}^\gamma(k_i) = k F_{t,i}^\beta(k_i) X_{t,i}^\gamma(k_i). \quad (12)$$

In the above equation, $k_F, k_X$ are proportionality constants of $F_{t,i}(k_i)$ and $X_{t,i}(k_i)$, respectively and $k = k_F k_X$. $\beta$ and $\gamma$ are positive, real-valued indices.

Here, we introduce a special notion of '*Priority*' ($Pri_{t,i}(k_i)$) of a sensor $i$ at any time slot $t$, based on the utility ($U_{t,i}(k_i)$) and laxity ($LX_{t,i}(k_i)$) its sample $k_i$ offers. It is defined as follows:

$$Pri_{t,i}(k_i) = \begin{cases} \frac{1}{U_{t,i}(k_i)\ X_{t+1,i}(k_i)} > 0, & for\ X_{t+1,i}(k_i) = 1 \\ \infty, & for\ X_{t+1,i}(k_i) = 0. \end{cases} \quad (13)$$

If the sample $k_i$ from the active sensor $i$ is critical at $t$, then its laxity $LX_{t,i}(k_i) = 0$ and $LX_{t+1,i}(k_i) = (LX_{t,i}(k_i) - 1) < 0$, if not served in the present slot. So, $X_{t+1,i}(k_i) = \mathcal{F}(LX_{t+1,i}(k_i)) = 0$ and $Pri_{t,i}(k_i) = \infty$ (by definition). On the other hand, if the sample from the active sensor $i$ is non-critical, then $LX_{t,i}(k_i) > 0$ and $LX_{t+1,i}(k_i) \geq 0$. So, $X_{t,i}(k_i) = X_{t+1,i}(k_i) = 1$ and $Pri_{t,i}(k_i) = \frac{1}{U_{t,i}(k_i)} = \frac{1}{kF_{t,i}^\beta(k_i)X_{t,i}^\gamma(k_i)} = \frac{1}{kF_{t,i}^\beta(k_i)}$ (from (11) and (12)). Moreover, from the calculation of the priority for the critical sensor considering the '*conflict avoidance*' mechanism, we can conclude that *a sensor having the critical sample has the highest priority among all the active sensors samples present at that particular slot $t$.*

Next, the expected weighted sum utility of information (EXWSUoI) is calculated as,

$$EXWSUoI = \frac{1}{TM} E\left[\sum_{t=1}^T \sum_{i=1}^M \{\alpha_i U_{t,i}(k_i) | \vec{I}\}\right]. \quad (14)$$

Here, $\vec{I} = [\overrightarrow{h_1(1)}, \overrightarrow{c(0)}, \overrightarrow{D(1)}]$ is the initial condition vector that represents the initial values of $h_{t,i}(k_i)$, $c_i(k_i - 1)$, and $D_i(k_i)$, respectively for $k_i = 1\ \forall i$. $\alpha_i = \alpha > 0\ \forall\ i \in [1, M]$ is the real-valued weight assigned to any sensor $i$ in the WSAN. From now onwards, the initial condition and packet index will be omitted for notation simplicity.

Now, we know that among $M$ sensors, only active sensors take part in the scheduling. Let, $S_t$ be the set of all the active sensors at the beginning of slot $t$. So, in order to maximize the utility of information, the objective function is considered as,

**Objective Function:** $O_T^{\pi^*} = \max_{\pi \in \Pi} E[O_T^{\pi}]$,

where, $O_T^{\pi} = \frac{\alpha}{TM}[\sum_{t=1}^{T} \sum_{i \in S_t} U_{t,i}]$.

*Finally, this paper aims to find an optimal work conserving, non-anticipative, online, dynamic priority based TDMA scheduling policy π, from the set of all the admissible policies Π, to maximize the objective function expected weighted sum utility of information (EXWSUoI).*

## IV. OPTIMALITY OF SCHEDULING POLICY

In this section, first, a greedy scheduling policy '*Deadline aware Highest Latency First (HLF-D)*' is proposed. Then, its optimality is analyzed in Theorem-I, followed by Lemma-I to III.

***Definition:*** *(i) At any time slot, if no critical sample is present in the system, HLF-D, a greedy scheduling policy, schedules the transmission of an unprocessed non-critical active sensor sample with the highest latency (or the lowest utility). (ii). The presence of a critical sample ensures that none but the critical sample will get service. (iii)[Tie-Breaking Condition] All ties are being broken arbitrarily.*

In other words, *greedy policy HLF-D schedules the sample with the highest priority first.*

The pseudocode for HLF-D is presented in algorithm 1.

***Lemma-I:*** *At any slot t, if a non-critical sensor sample i has higher utilization than that of other non-critical sample j, then i is going to provide higher utilization than j in the next slot, as well, if both of them are not served at slot t.*

**Proof:** At slot $t$, the utilization $U_{t,i}$ of a non-critical sensor sample $i$ is higher than the utilization $U_{t,j}$ of any other non-critical sample $j$. So, $U_{t,i} \leq U_{t,j}$. We have to prove that, in the next slot $t+1$, if not served, $U_{t+1,i} \leq U_{t+1,j}$.

If non-critical samples $i$ and $j$ are not served at slot $t$, in the next slot $t+1$, both of their latency ($L_{t,i}$ and $L_{t,j}$) will increase by +1 and laxity ($LX_{t,i}$ and $LX_{t,j}$) will reduce by (-1). So, their freshness will change from $F_{t,i} = \frac{1}{L_{t,i}+P_i}$ to $F_{t+1,i} = \frac{1}{L_{t,i}+1+P_i} = \frac{1}{L_{t,i}+1+1}$ and from $F_{t,j} = \frac{1}{L_{t,j}+P_j}$ to $F_{t+1,j} = \frac{1}{L_{t,j}+1+P_j} = \frac{1}{L_{t,j}+1+1}$, respectively. [Note: the processing time $P_i = P_j = 1$ always for this system model.] As both the samples are non-critical, their laxity at slot $t$ are, $LX_{t,i} > 0$ and $LX_{t,j} > 0$ and laxity at slot $t+1$ will become, $LX_{t+1,i} \geq 0$ and $LX_{t+1,j} \geq 0$. So, for both $t$ and $t+1$, $X_{t,i} = X_{t+1,i} = 1$. This gives, $U_{t,i} = kF_{t,i}^{\beta}X_{t,i}^{\gamma} = kF_{t,i}^{\beta}$, $U_{t+1,i} = kF_{t+1,i}^{\beta}X_{t+1,i}^{\gamma} = kF_{t+1,i}^{\beta}$ and $U_{t,j} = kF_{t,j}^{\beta}X_{t,j}^{\gamma} = kF_{t,j}^{\beta}$, $U_{t+1,j} = kF_{t+1,j}^{\beta}X_{t+1,j}^{\gamma} = kF_{t+1,j}^{\beta}$. Now, from the statement of the Lemma-I,

$$U_{t,i} \leq U_{t,j}$$

or, $$kF_{t,i}^{\beta} \leq kF_{t,j}^{\beta}$$

**Algorithm 1:** Age evolution for active-inactive-out-of-service sensors using HLF-D policy.

/* Initialization */
1  input $\overrightarrow{h_1(1)}, \overrightarrow{c(0)}, \overrightarrow{D(1)}, T, M$
2  $t \leftarrow 1$, $k \leftarrow 1$ for all sensors $i \in [1, M]$
3  while ($t \leq T$)
4    for all $i \in [1, M]$, do
/* Finding active-inactive-out-of-service sensors at slot $t$ */
5      if $h_{t,i}(k_i) < c_i(k_i - 1) + 1$ then
/* Age evolution of inactive sensors at slot $t$ */
6        $h_{t+1,i}(k_i) = h_{t,i}(k_i) + 1$
7      elseif
8        $h_{t,i}(k_i) > D_i(k_i)$ then,
/* Age evolution of out-of-service sensors at slot $t$ */
9        $h_{t+1,i}(k_i) = h_{t,i}(k_i) + 1$
10     else
/* Store active sensors in a set $S_t^{HLF-D}$ */
11       $S_t^{HLF-D} \leftarrow i$
12     end
13   end
   Only active sensors take part in the scheduling
14   for all $i \in S_t^{HLF-D}$, do
/* Calculate priority for each active sensor sample at slot $t$ */
A critical sensor has the highest priority among all sensors. In the absence of any critical sensor, among all non-critical sensors, sensor, having a sample with the highest latency, has the highest priority.
15     $Pri_{t,i}(k_i) = \frac{1}{U_{t,i}(k_i)X_{t+1,i}(k_i)}$

/* Scheduling decision in slot $t$ */
HLF-D schedules either the critical sample or the unprocessed active sensor sample having the highest latency (in absence of critical sample). From the priorities of active sensors (in line 15) it can be seen that HLF-D basically serves the sensor which has the highest priority .
16     $\omega^{t(HLF-D)} = \arg\max_{i \in S_t^{HLF-D}} \{Pri_{t,i}(k_i)\}$

/* Post decision age evolution of active sensor at slot $t$ */
The scheduled sensor sample is processed in slot $t$ according to the channel state.
17     if channel is ON and $i == \omega^{t(HLF-D)}$ then,
18       $h_{t+1,i} = 1$, $c_i(k_i - 1) \leftarrow c_i(k_i), k_i \leftarrow k_i + 1$
19     else
20       $h_{t+1,i}(k_i) = h_{t,i}(k_i) + 1$
21     end
22   end
23   $t \leftarrow t + 1$
24 end

or, $\quad F_{t,i} \leq F_{t,j}$

or, $\quad \frac{1}{L_{t,i}+P_i} \leq \frac{1}{L_{t,j}+P_j}$

or, $\quad L_{t,i} + P_i \geq L_{t,j} + P_j$

or, $\quad L_{t,i} \geq L_{t,j} \quad$ (As, $P_i = P_j = 1$)

or, $\quad \frac{1}{L_{t,i}+1+1} \leq \frac{1}{L_{t,j}+1+1}$

or, $\quad F_{t+1,i} \leq F_{t+1,j}$

or, $\quad kF_{t+1,i}^{\beta} \leq kF_{t+1,j}^{\beta}$

or, $\quad U_{t+1,i} \leq U_{t+1,j} \quad$ (∴ Proved) ∎

**Lemma-II:** Number of elements present in $S_t^{HLF-D}$ is not lesser than that of $S_t^{\pi}$.

It is known that HLF-D and $\pi$ both started from the same initial condition. HLF-D always processes the critical sample or, in the absence of any critical packet, serves the packet with the highest latency incurred so far. The same is not true for $\pi$. $\pi$ serves a sample with lower latency or any sample other than the critical one. So, critical packets may get dropped in $\pi$ and the number of active sensors in $S_t^{\pi}$ maybe less than that in $S_t^{HLF-D}$. ∎

**Lemma-III.** At any time slot, the critical sample present in the active set $S_t^{\pi}$ has either same or lower utilization than the critical sample present in the active set $S_t^{HLF-D}$.

From Lemma-II, it is known that starting from the same initial condition, HLF-D always serves either the critical sample or the sample having the highest latency (in the absence of any critical sensor). Other active samples with lower latencies are inherited to the active set in the next slot. But this is not the case in $\pi$. $\pi$ serves either sample with lower latency or any sample other than the critical one, and this critical packet is dropped. As a result, one of the two cases may occur.

i. Same critical sample is present in both $S_t^{\pi}$ and $S_t^{HLF-D}$.
ii. Different critical samples are present in $S_t^{\pi}$ and $S_t^{HLF-D}$.

*Case i* is quite obvious as both $S_t^{HLF-D}$ and $S_t^{\pi}$ initially starts from the same sample space. *Case ii* can occur only when any common non-critical sample having the highest latency is not served by $\pi$ but, it is served by HLF-D in any prior slot. This leftover sample is inherited to the future slots and is present in $S_t^{\pi}$ as the critical sample having the hard deadline. So, it is supposed to have higher latency (or, lower utilization) than the critical samples present in $S_t^{HLF-D}$. ∎

**Theorem-I (Optimality of HLF-D algorithm):** For any symmetric IWSAN network with an unreliable time-shared channel from the sensor to the processor, among the class of admissible policies HLF-D attains the maximum expected weighted sum utility of information ($O_T^{\pi^*}$) for sensor samples with deadlines.

**Proof:** To prove the optimality of HLF-D scheduling policy, the value of the objective function obtained by HLF-D and any other admissible policy $\pi \in \Pi$ are compared.

In the objective function, let $\sum_{i \in S_t} U_{t,i} = V_t$. As, $\alpha, T, M$ are constants, hence, if it can be proved that $V_t^{HLF-D} \geq V_t^{\pi} \forall t \in [1,T]$, then it is sufficient to state that $O_T^{\pi^*} = E[O_T^{HLF-D}] \geq E[O_T^{\pi}] \forall \pi \in \Pi$.

Now, let us denote the elements in $S_t$ as $x_a$ for $a$ in the range [1, M]. So, $S_t = \{x_a\}_t$ for $a = 1 \ldots R^{(t)}$ be the set of active sensors at the beginning of slot $t$. $R^{(t)}$ is the number of elements present in the active set i.e. $|S_t| = R^{(t)} \leq M$. $S_t' = \{e_{a'}\}_t$ for $a' = 1 \ldots R^{(t)}$ is the rearranged set $S_t$, arranged in decreasing order of priority of the constituents. However, the priority of samples from $a^{th}$ and $a'^{th}$ active sensors in the active sensor set $S_t$ and rearranged active sensor set $S_t'$ are denoted as $Pri_{t,a}$ and $Pri_{t,a'}$, respectively. So, in the objective function, the sum of utility of the active sensor samples becomes $V_t = \sum_{i \in S_t} U_{t,i} = \sum_{x_a \in S_t} U_{t,a} = \sum_{e_{a'} \in S_t'} U_{t,a'}$. Here, $U_{t,a}$ and $U_{t,a'}$ are the utilizations of samples from $a^{th}$ and $a'^{th}$ active sensors in sets $S_t$ and $S_t'$, respectively.

Next, to prove $V_t^{HLF-D} \geq V_t^{\pi} \forall t \in [1,T]$ which in turn, proves $O_T^{\pi^*} = E[O_T^{HLF-D}] \geq E[O_T^{\pi}] \forall \pi \in \Pi$, without loss of generality, we finally try to prove that $U_{t,a'}^{HLF-D} \geq U_{t,a'}^{\pi} \forall e_{a'} \in S_t'$ and $t \in [1,T]$.

We are using the **Forward Induction method** for this proof.

Let's say, at any time slot $t$, the active sensor from $S_t'$, scheduled for transmission, is denoted by $\omega^t$. If the channel is 'ON,' the sample from the active sensor $\omega^t$ is processed successfully during this slot. However, if any critical sample is present in $S_t'$ at slot $t$, then this sample is going to be dropped at the end of this slot, if not scheduled for transmission. Active sensor from which critical sample is dropped is denoted by $drop^t$. In the next slot $t+1$, active set $S_{t+1} = S_t' \setminus \{\omega^t, drop^t\} \cup N_{t+1}$ where $N_{t+1} = \{y_n\}_{t+1}$ for $n = 1 \ldots r^{(t+1)}$ is the set of newly active sensors at the beginning of slot $t+1$. We put a 0 for the empty place of $drop^t$ at the beginning of the set $S_{t+1}$. This zero-padding implicates that, according to our system model, the particular flow-line $drop^t$ will go out of service, and its corresponding actuation task will be handled by some default command from the controller. However, this sensor will stay in the active set as a dummy element with a sample having $utility = 0$ and $laxity < 0$. But, this out-of-service sensor does not take part in the scheduling. Therefore, the number of elements in $S_{t+1}$ becomes $R^{(t+1)} = [max(0, (R^{(t)} - 1)) + r^{(t+1)}] \leq M$.

Now, if any active sensor with critical sample ($utility > 0$, $laxity = 0$ and $priority = \infty$) pops up in this slot $t+1$, we place that sensor in $S_{t+1}'$ before those dummy zeros followed by the active sensors with non-critical samples in the decreasing order of their priority.

*Assumptions:*
  I. For comparing the policies HLF-D and $\pi$, it is imperative that the channel condition, the number of newly active sensors, and the deadlines of their sensor samples at the beginning of each slot remain independent of the underlying scheduling policy and does not change during one slot.
  II. In this proof sensor and sensor sample, both the terms can be used interchangeably. This is because the samples are present within the sensor itself.

**Base case:** Initial conditions are the same for both the policies. So, the set of active sensors $S_t$ is the same for both HLF-D and $\pi$ at slot $t = 1$. This yields $U_{1,a'}^{HLF-D} = U_{1,a'}^{\pi} = U_{1,a'}$ $\forall e_{a'} \in S_1'$.

**Inductive step:** For any time slot $t$, it is assumed that, $U_{t,a'}^{HLF-D} \geq U_{t,a'}^{\pi}$ $\forall e_{a'} \in S_t'$. It is required to prove that, $U_{t+1,a'}^{HLF-D} \geq U_{t+1,a'}^{\pi}$ $\forall e_{a'} \in S_{t+1}'$.

At any slot $t$, active sets obtained from two policies HLF-D and $\pi$ are as follows:

$$S_t'^{HLF-D} = \{e_1, e_2, \ldots e_i, \ldots, e_j, \ldots, e_k, \ldots, e_{R^{(t)}}\}_t^{HLF-D}$$

$$S_t'^{\pi} = \{e_1, e_2, \ldots, e_i, \ldots, e_j, \ldots, e_k, \ldots, e_{R^{(t)}}\}_t^{\pi}. \quad (15)$$

Now, one of the four distinct cases may happen:
1. Critical sensors are present in both $S_t'^{HLF-D}$ and $S_t'^{\pi}$.
2. Critical sensor is present only in $S_t'^{HLF-D}$ but not in $S_t'^{\pi}$.
3. Critical sensor is present only in $S_t'^{\pi}$ but not in $S_t'^{HLF-D}$.
4. No critical sensor is present in either $S_t'^{HLF-D}$ or $S_t'^{\pi}$.

According to HLF-D, the active sensor, having the highest priority, is scheduled for processing. This follows, $\omega^{t(HLF-D)} = arg \max_{x_a \in S_t^{HLF-D}} \{Pri_{t,a}^{HLF-D}\} = \{e_1\}_t^{HLF-D}$.
Whereas, in policy $\pi$ any element other than the active sensor with the highest priority is scheduled for processing. This gives, $\omega^{t(\pi)} = \{e_k\}_t^{\pi} \neq arg \max_{x_a \in S_t^{\pi}} \{Pri_{t,a}^{\pi}\} = \{e_1\}_t^{\pi}$.
Here, $1 < k \leq R^{(t)}$.

Now, we compare HLF-D and $\pi$ for the aforementioned four cases one by one.

According to Lemma II, the number of elements present in $S_t^{\pi}$ is lesser than or equal to the elements present in $S_t^{HLF-D}$. Thus, for comparing HLF-D and $\pi$, at first, the number of elements in $S_t^{\pi}$ is made the same as that in $S_t^{HLF-D}$ by adding dummy zeros at the beginning of $S_t^{\pi}$.

**A. Case 1.** We know that the critical has the highest priority in an active set. So, in $S_t'^{HLF-D}$ and $S_t'^{\pi}$, $\{e_1\}_t^{HLF-D}$ and $\{e_1\}_t^{\pi}$ are the sensors with critical samples, respectively. From earlier discussions, $drop^{t(HLF-D)} = \emptyset$ and $\omega^{t(HLFD)} = \{e_1\}_t^{HLF-D}$. So, in HLF-D,

$$S_t'^{HLF-D} = \{\cancel{e_1}, e_2 \ldots e_k \ldots e_{R^{(t)}}\}_t^{HLF-D}$$

$$S_{t+1}^{HLF-D} = \{e_2 \ldots e_k \ldots e_{R^{(t)}}\}_t^{HLF-D} \cup \{y_1 \ldots y_{r^{(t+1)}}\}_{t+1}^{HLF-D}$$

$$= \{x_1 \ldots x_{k-1} \ldots x_{R^{(t)}-1}, x_{R^{(t)}} \ldots x_{R^{(t+1)}}\}_{t+1}^{HLF-D}$$

$$= \{S_t'^{HLF-D} \setminus \{e_1\}_t^{HLF-D}\} \cup N_{t+1}^{HLF-D}. \quad (16)$$

From (16), the relation between $S_{t+1}^{HLF-D}$ and $S_t'^{HLF-D}$ can be obtained as:

$$\{x_a\}_{t+1}^{HLF-D} = \{e_{a'+1}\}_t^{HLF-D} \quad for\ 1 \leq a = a' < R^{(t)}$$

$$= \{y_n\}_{t+1}^{HLF-D} \quad for\ n = 1\ to\ r^{(t+1)}\ and$$
$$R^{(t)} \leq a \leq R^{(t+1)}. \quad (17)$$

Here, all the sensors $\{e_{a'+1}\}_t^{HLF-D}$ for $1 \leq a = a' < R^{(t)}$ are having non-critical samples at slot $t$ and $\{y_n\}_{t+1}^{HLF-D}$ for $n = 1$ to $r^{(t+1)}$ has just arrived at slot $t + 1$. So, the utilization of the elements in $S_{t+1}^{HLF-D}$ are,

$$U_{t+1,a}^{HLF-D} = U_{t+1,\{a'+1\}_t}^{HLF-D} \quad for\ 1 \leq a = a' < R^{(t)}$$

$$= 1 \quad for\ R^{(t)} \leq a \leq R^{(t+1)}. \quad (18)$$

Here, $U_{t+1,\{a'\}_t}^{HLF-D}$ implies utilization of $a'$ element from $S_t'^{HLF-D}$ in the next slot $t + 1$.

On the other hand, $drop^{t(\pi)} = \{e_1\}_t^{\pi}$ and $\omega^{t(\pi)} = \{e_k\}_t^{\pi} \neq \{e_1\}_t^{\pi}$. $\{e_1\}_t^{\pi}$ & $\{e_k\}_t^{\pi}$ are active sensors having samples with non-zero utility. So in $\pi$,

$$S_t'^{\pi} = \{\cancel{e_1}, e_2 \ldots e_{k-1}, \cancel{e_k}, e_{k+1} \ldots e_{R^{(t)}}\}_t^{\pi}$$

$$S_{t+1}^{\pi} = \{0, e_2 \ldots e_{k-1}, e_{k+1} \ldots e_{R^{(t)}}\}_t^{\pi} \cup \{y_1 \ldots y_{r^{(t+1)}}\}_{t+1}^{\pi}$$

$$= \{x_1, x_2 \ldots x_{k-1}, x_k \ldots x_{R^{(t)}-1}, x_{R^{(t)}} \ldots x_{R^{(t+1)}}\}_{t+1}^{\pi}$$

$$= \{S_t'^{\pi} \setminus \{e_1, e_k\}_t^{\pi}\} \cup N_{t+1}^{\pi}. \quad (19)$$

From (19), the relation between $S_{t+1}^{\pi}$ and $S_t'^{\pi}$,

$$\{x_a\}_{t+1}^{\pi} = 0 \quad for\ a = 1.$$

$$= \{e_{a'}\}_t^{\pi} \quad for\ 1 < a = a' < k.$$

$$= \{e_{a'+1}\}_t^{\pi} \quad for\ k \leq a = a' < R^{(t)}.$$

$$= \{y_n\}_{t+1}^{\pi} \quad for\ n = 1\ to\ r^{(t+1)}\ and$$
$$R^{(t)} \leq a \leq R^{(t+1)}. \quad (20)$$

Other than $\{e_1\}_t^\pi$ all elements in $S_t'^\pi$ are non-critical. So, the utility of the elements in $S_{t+1}^\pi$ are,

$$U_{t+1,a}^\pi = 0 \qquad for\ a = a' = 1.$$
$$= U_{t+1,\{a'\}_t}^\pi \qquad for\ 1 < a = a' < k.$$
$$= U_{t+1,\{a'+1\}_t}^\pi \qquad for\ k \le a = a' < R^{(t)}.$$
$$= 1 \qquad for\ n = 1\ to\ r^{(t+1)},$$
$$R^{(t)} \le a \le R^{(t+1)}. \quad (21)$$

Now comparing (18) and (21) it is observed that,

**for $a = a' = 1$:** $U_{t+1,a}^{HLF-D} = U_{t+1,\{a'+1\}_t}^{HLF-D} > 0$ and $U_{t+1,a}^\pi = 0$. So, $U_{t+1,a}^{HLF-D} \ge U_{t+1,a}^\pi$.

**for $1 < a = a' < k$:** In this segment, all the sensors are non-critical in both $S_t'^{HLF-D}$ and $S_t'^\pi$ and $Pri_{t,a'} > Pri_{t,a'+1}$. So, $U_{t,a'} \le U_{t,a'+1}$. Combining this with our initial assumption of the induction step, it is obtained that $U_{t,a'+1}^{HLF-D} \ge U_{t,a'}^{HLF-D} \ge U_{t,a'}^\pi$. Now, from Lemma-I, we can say that if $U_{t,a'+1}^{HLF-D} \ge U_{t,a'}^\pi$ then $U_{t+1,\{a'+1\}_t}^{HLF-D} \ge U_{t+1,\{a'\}_t}^\pi$, as well. This proves, $U_{t+1,a}^{HLF-D} \ge U_{t+1,a}^\pi$.

**for $k \le a = a' < R^{(t)}$:** From the initial assumption, we can say that $U_{t,a'+1}^{HLF-D} \ge U_{t,a'+1}^\pi$. All the sensors are having non-critical samples in this segment, too, in both $S_t'^{HLF-D}$ and $S_t'^\pi$. So, from Lemma-I, we can show that $U_{t+1,\{a'+1\}_t}^{HLF-D} \ge U_{t+1,\{a'+1\}_t}^\pi$ or in other words, $U_{t+1,a}^{HLF-D} \ge U_{t+1,a}^\pi$.

**for $R^{(t)} \le a \le R^{(t+1)}$:** $U_{t+1,a}^{HLF-D} = U_{t+1,a}^\pi = 1$.

From the above comparisons, it can be concluded that $U_{t+1,a}^{HLF-D} \ge U_{t+1,a}^\pi\ \forall x_a \in S_{t+1}$. Again, the four sub-cases, similar to the abovementioned cases (1-4), may occur in slot $t+1$ depending on the presence of a critical sensor in any active set. They are analyzed case by case to complete the induction.

**A.1. Subcase 1.1:** Let, $\{x_I\}_{t+1}^{HLF-D}$ and $\{x_J\}_{t+1}^\pi$ are having critical samples in $S_{t+1}^{HLF-D}$ and $S_{t+1}^\pi$, respectively. Here $I$ and $J$ both are smaller than $R^{(t+1)}$. So, the relation between $S_{t+1}^{HLF-D}$ and $S_{t+1}'^{HLF-D}$ are as follows,

$$S_{t+1}^{HLF-D} = \{x_1, x_2 \dots x_{I-1}, \boxed{x_I}, x_{I+1} \dots x_{R^{(t+1)}}\}_{t+1}^{HLF-D}$$
$$S_{t+1}'^{HLF-D} = \{\boxed{x_I}, x_1, x_2 \dots x_{I-1}, x_{I+1} \dots x_{R^{(t+1)}}\}_{t+1}^{HLF-D}$$
$$= \{\boxed{e_1}, e_2, e_3 \dots e_I, e_{I+1}, \dots e_{R^{(t+1)}}\}_{t+1}^{HLF-D}. \quad (22)$$

This gives,

$$\{e_{a'}\}_{t+1}^{HLF-D} = \{x_I\}_{t+1}^{HLF-D} \qquad for\ a' = 1$$
$$= \{x_{a-1}\}_{t+1}^{HLF-D} \quad for\ 1 < a = a' \le I$$
$$= \{x_a\}_{t+1}^{HLF-D} for\ I < a = a' \le R^{(t+1)}. \quad (23)$$
$$U_{t+1,a'}^{HLF-D} = U_{t+1,I}^{HLF-D} \qquad for\ a' = 1$$
$$= U_{t+1,a-1}^{HLF-D} \qquad for\ 1 < a = a' \le I$$
$$= U_{t+1,a}^{HLF-D} \quad for\ I < a = a' \le R^{(t+1)}. \quad (24)$$

Similarly in $\pi$,

$$S_{t+1}^\pi = \{x_1, x_2 \dots x_{J-1}, \boxed{x_J}, x_{J+1} \dots x_{R^{(t+1)}}\}_{t+1}^\pi$$
$$S_{t+1}'^\pi = \{\boxed{x_J}, x_1, x_2 \dots x_{J-1}, x_{J+1} \dots x_{R^{(t+1)}}\}_{t+1}^\pi$$
$$= \{\boxed{e_1}, e_2, e_3 \dots e_J, e_{J+1}, \dots e_{R^{(t+1)}}\}_{t+1}^\pi. \quad (25)$$

From (25) we get,

$$\{e_{a'}\}_{t+1}^\pi = \{x_J\}_{t+1}^\pi \qquad for\ a' = 1$$
$$= \{x_{a-1}\}_{t+1}^\pi \qquad for\ 1 < a = a' \le J$$
$$= \{x_a\}_{t+1}^\pi \quad for\ J < a = a' \le R^{(t+1)}. \quad (26)$$
$$U_{t+1,a'}^\pi = U_{t+1,J}^\pi \qquad for\ a' = 1$$
$$= U_{t+1,a-1}^\pi \qquad for\ 1 < a = a' \le J$$
$$= U_{t+1,a}^\pi \quad for\ J < a = a' \le R^{(t+1)}. \quad (27)$$

Now comparing (24) and (27),

**for $a' = 1$:** $U_{t+1,a'}^{HLF-D} = U_{t+1,I}^{HLF-D}$ and $U_{t+1,a'}^\pi = U_{t+1,J}^\pi$.

From Lemma III, it is known that at any time slot, the critical sample present in the active set for $\pi$, has the same or lower utilization than the critical sample present in the active set for HLF-D. Therefore, comparing the utilization of critical samples in $S_{t+1}^{HLF-D}$ and $S_{t+1}^\pi$, we get that, $U_{t+1,I}^{HLF-D} \ge U_{t+1,J}^\pi$. This proves $U_{t+1,a'}^{HLF-D} \ge U_{t+1,a'}^\pi\ for\ a' = 1$.

**for $a = a' > 1$:** If $I \le J$, from (24) and (27) it can be seen directly that $U_{t+1,a'}^{HLF-D} \ge U_{t+1,a'}^\pi$. Otherwise, we look back to the elements in previous sets $S_t'^{HLF-D}$ in (18) and $S_t'^\pi$ in (21) to replace values of $U_{t+1,a}^{HLF-D}$ and $U_{t+1,a}^\pi$ in (24) and (27), respectively. From backtracking, it can be concluded that, $U_{t+1,a'}^{HLF-D} \ge U_{t+1,a'}^\pi$ for $I > J$ iff $I < k$. Next, we show that if $I > J$, then it naturally follows $I < k$.

If the same critical sensor is present both in $S_{t+1}^{HLF-D}$ and $S_{t+1}^\pi$ and its position be $I$ in $S_{t+1}^{HLF-D}$, then in $S_{t+1}^\pi$, it must be in the same position or after $I$ (due to initial 0 padding). That means $I \le J$. But $I > J$ means two completely different sensors are critical in $S_{t+1}^{HLF-D}$ and $S_{t+1}^\pi$. This can happen only when the element from $S_{t+1}^\pi$ same as the $I^{th}$ sample in $S_{t+1}^{HLF-D}$ is already served by $\pi$ at the previous time slot $t$ or any earlier slot. Let say, the $k^{th}$ sensor in $S_t'^\pi$ is same as the $I^{th}$ sensor in $S_{t+1}^{HLF-D}$.

Now, at any slot $t$, the first element from $S'^{HLF-D}_t$ is being served. Other elements are shifted one position left, and they are inherited to the active set $S^{HLF-D}_{t+1}$ in the next slot. So, $I^{th}$ sensor in $S^{HLF-D}_{t+1}$ was in $(I+1)^{th}$ position in $S'^{HLF-D}_t$. Therefore, between $S'^{\pi}_t$ and $S'^{HLF-D}_t$, $k \geq (I+1)$, or in other words, $I < k$ always.

For the limited scope of discussion in this paper, only one case and its one subcase have been analyzed thoroughly considering the channel is *ON*. When the channel is *OFF*, it affects both the policies HLF-D and $\pi$. Thus, the status of the active sensors will not change only their individual utility will degrade with time. Analysis for the other cases and their all possible subcases proceeds in the same manner, and finally, it can be proved that if $U^{HLF-D}_{t,a'} \geq U^{\pi}_{t,a'}$, $\forall e_{a'} \in S'_t$ at any slot $t$, in the next slot $t+1$ $U^{HLF-D}_{t+1,a'} \geq U^{\pi}_{t+1,a'}$ $\forall e_{a'} \in S'_{t+1}$ as well (**Induction complete**). ∎

## V. RESULTS AND DISCUSSION

This section simulates in MATLAB the performance of HLF-D in terms of EXWSUoI, $\overline{AoI}$, AL and RMS jitter (EXWSUoI has the unit (No of time slots)$^{-1}$. Rest all the parameters are expressed in terms of the number of time slots) for symmetric IWSAN network and compare the results with those of Highest Latency First (HLF) policy [17] and other traditional algorithms viz. Earliest Deadline First (EDF) and Least Laxity First (LLF). All the algorithms, as mentioned above, including our proposed one, have the complexity of order $O(n)$. [detailed proof is given in Appendix A].

For simulation, we have considered a symmetric system model of M sensor-actuator pairs connected to a centralized controller through an unreliable TDMA shared channel with '*ON*' probability $p_i = p$. The controller is serving one packet per slot, and the 'age' of information is evolving as described in the system model of the paper. Our aim is to maximize the objective function EXWSUoI, which, in turn, is expected to minimize the average age $\overline{AoI}$ (eq. 7), average latency $AL$ (eq. 8), and RMS jitter $\overline{JT}$ (eq. 9) of the system, too. It is assumed that the channel condition, the number of newly active sensors, and deadlines of their sensor samples at the beginning of each slot remain independent of the underlying scheduling policy and does not change during one slot.

Parameters along with their values, used for the simulations, are listed in Table I. Initially, time slot $t$ and packet index $k_i \forall i$ is considered to be 1. For all sensors $i \in [1, M]$, the initial age is considered to be $1 \leq h_{1,i}(1) < D_i(1) = 1 + c_i(0) + d_i(1)$ slots.

Fig. 4. plots the comparison of EXWSUoI values varying with T for different algorithms. From this plot, it can be seen that the expected utilization of information is always maximum for HLF-D for any value of $T$. For policies like HLF and EDF, EXWSUoI drops drastically with the increasing value of $T$.

Fig. 5. shows the comparisons for average AoI, Average latency for the sensor samples, and RMS jitter in the actuators for different algorithms with respect to varying $T$. It can be noticed that for a sufficiently lower value of T, the

TABLE I
SIMULATION PARAMETERS

| Parameters | Symbols | Values Assigned |
|---|---|---|
| Number of flow-line (sensor-actuator pair) [17], [21] | M | 16 |
| The weight assigned to each sensor | $\alpha_i = \alpha > 0$ $\forall i \in M$ | 1 |
| Time-shared transmission channel reliability | $p_i = p$ $\forall i \in M$ | 0.8 |
| Time slot (According to WirelessHart standard [20]) | $t$ | 10 ms |
| Initial useful age [21] | $\vec{c(0)}$ | [1,25] |
| Initial deadline [22] | $\vec{D(1)}$ | [1,20] |
| Real Indices | $\beta > 0$, $\gamma > 0$ | 1 |
| Proportionality constants | $k_F, k_X$ | 1 |

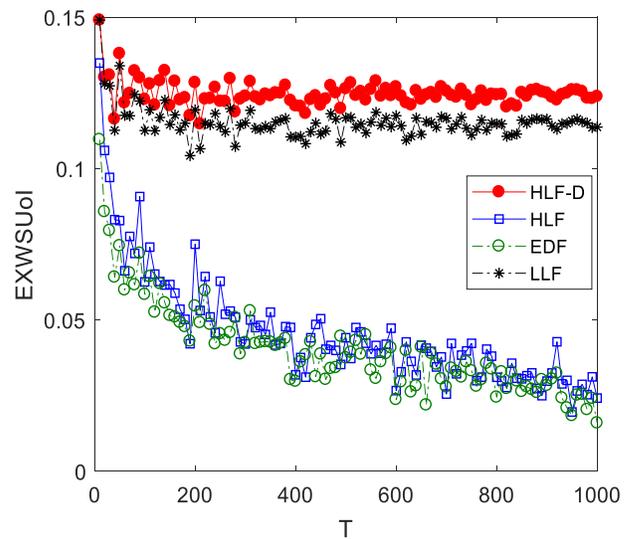

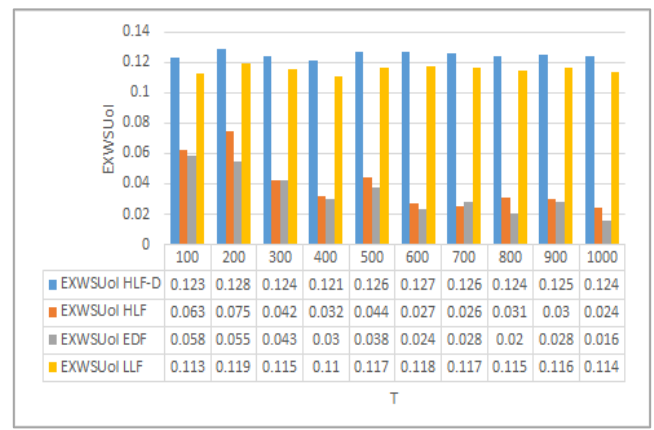

Fig. 4. Comparisons of EXWSUoI vs T for different algorithms.

performances of the other three algorithms are comparable to HLF-D. This is because the effect of packet deadlines is not significant for a small value of $T$. But, as time passes, the dominance of our proposed policy HLF-D over other algorithms in deadline-aware scheduling becomes prominent.

Fig. 6 plots the mean value obtained of the average age of information, latency, and RMS jitter of the sensor-actuator network over $T$ for $T = 100, 200, 300, ..., 1000$. It can be clearly seen that our proposed HLF-D algorithm provides the lowest mean values for all of the preferred metrics than those of the other scheduling policies. For all the parameters in Fig 4-6, HLF and EDF provide much inferior performances than the other two algorithms.

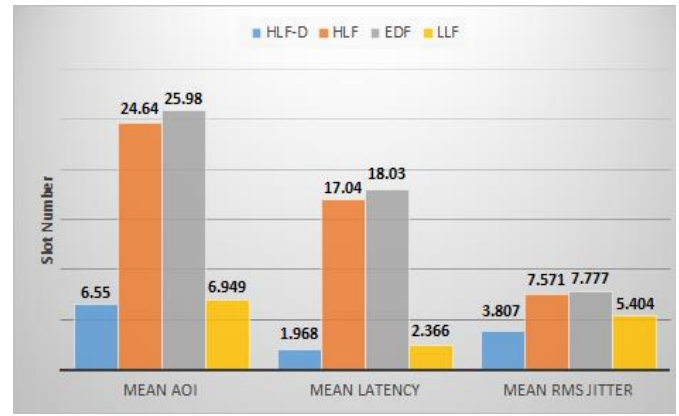

Fig. 6. Comparisons of mean value obtained of age, latency, and RMS jitter for different algorithms by averaging over T.

Though a smaller value of average age should be maintained to keep the information fresh for accurate decision making in the cyber systems, but the coexistence of cyber and physical systems in CPSs make the problem of maximizing data freshness more challenging. QoS (in terms of latency, delay, jitter, etc.) in the actuation tasks is also a major factor here that should also be taken care of besides accurate decision making. Accurate decision making and better QoS both are essential in remote monitoring and closed-loop, networked control of the cyber-physical system (CPS). Reference [17] proves that minimizing the latency content, involved in the age, is more effective for the thoughtful trade-off between information freshness maximization and jitter minimization in CPSs. Moreover, ICPS is a time-critical real-time system. Here information exchange should be done as timely as possible to guarantee real-time responses. Whenever a packet misses its deadline, it is dropped, causing production loss, accidents, or some fatal consequences. Therefore, minimizing packet loss has the utmost importance, too, in addition to latency minimization.

*The algorithm HLF proposed in [17] provides optimality in terms of data freshness and network performance in ICPS only when the packet deadline is not being considered. But, the deadline is one of the most important features associated with packet transmission whenever real-time communication is concerned. If a packet fails to meet its deadline, some severe consequences may take place due to unwanted packet loss. For addressing this challenge, a special deadline-aware scheduling policy is required, and our proposed greedy scheduling policy deadline-aware highest latency first (HLF-D) serves this purpose successfully.*

*HLF-D maximizes the proposed metric expected utility value of all the time-critical information in IWSAN. This can be done by minimizing the average latency of the sensor samples while attaining the minimum packet drops. Thus, by*

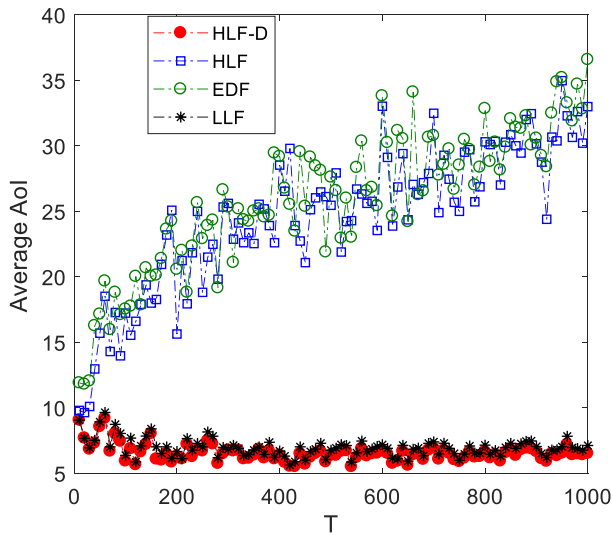

(a)

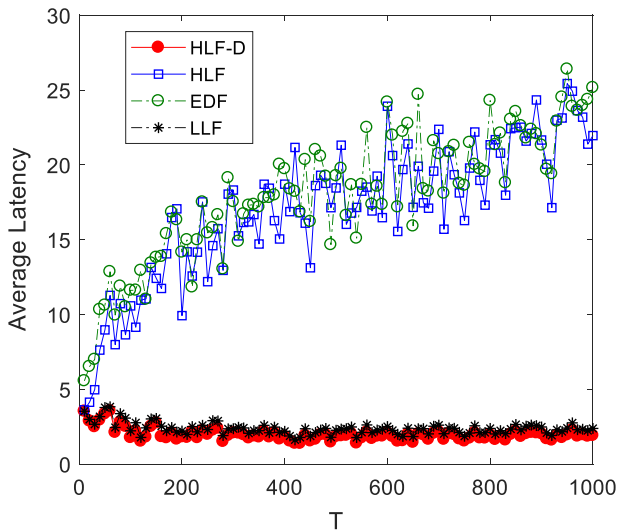

(b)

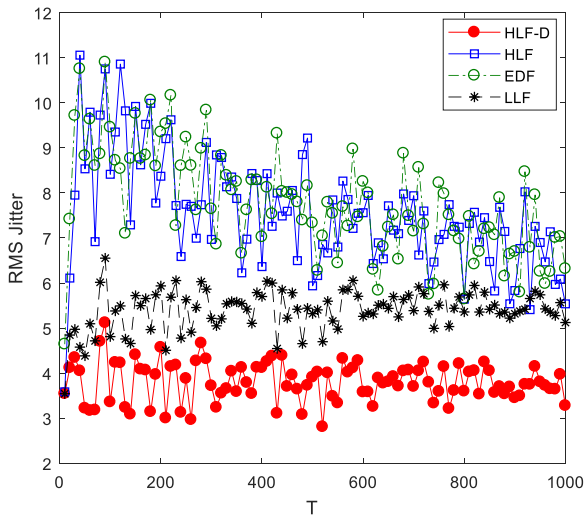

(c)

Fig. 5. Comparisons of (a) average age, (b) average latency, and (c) RMS jitter for different algorithms.

maximizing the utility, in turn, HLF-D satisfies the objective of attaining maximum information freshness. Moreover, algorithm HLF-D also provides sufficiently lower value for parameters like age, latency, and RMS jitter of the actuation tasks, as compared to those obtained by the other traditional scheduling algorithms. The performance of the system in terms of these parameters are of unavoidable concern for providing the satisfactory QoS performance of the IWSANs.

So, one may aptly conclude that HLF-D is the efficient deadline-aware scheduling strategy for optimizing the freshness of information and QoS in ICPS.

## VI. CONCLUSION

This paper analyzed the age-based information freshness in symmetric IWSAN application. One deadline-aware, dynamic priority-based greedy sensor scheduling algorithm '*Deadline-aware highest latency first*' was proposed for that purpose. Its optimality has been proved analytically in terms of the utility value of information content. Moreover, its effects on system performance in terms of mean age, latency, and RMS jitter were also compared with that of the highest latency first and other traditional scheduling algorithms by extensive simulations. From this paper, it can be concluded that to guarantee a real-time response and data freshness simultaneously, our proposed algorithm is the most effective packet scheduling scheme for symmetric IWSAN with a centralized controller, forming a CPS suitable for industrial applications.

## APPENDIX A
## TIME COMPLEXITY ANALYSIS FOR HLF-D POLICY

From Algorithm I, it can be said that the HLF-D runs in the form as given in Fig. 7. There are two "*for*" loops in a row. The first "*for loop*" consists of an "*if-then-elseif-then-else*" statement, and the second "*for loop*" consists of an "*if-then-else*" statement and both the loops are free from any kind of inner loops. Therefore, the worst-case runtime [23] of the algorithm will be, $T(n) = n*O(1) + n*O(1) = O(n)$. This can also be written as $O(\max(n,n)) = O(n)$. Therefore, it is proved that for the HLF-D algorithm, the time complexity is $O(n)$.

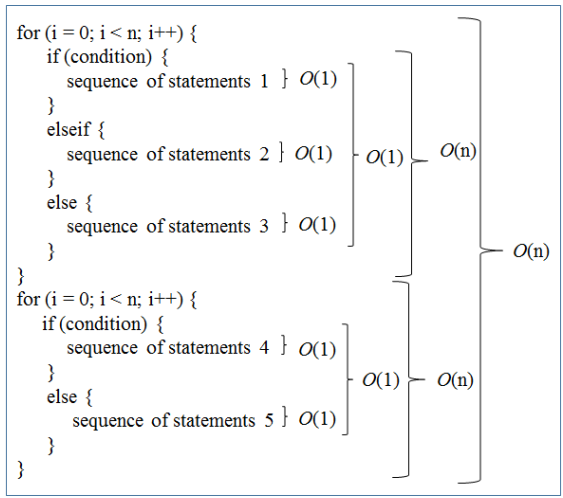

Fig. 7. Time complexity calculation of HLF-D algorithm.

## APPENDIX B
## LIST OF PARAMETERS

Table II lists the important notations used in this paper.

TABLE II
List of Notations

| Notations | Descriptions |
|---|---|
| M | Number of Sensor-actuator pairs in symmetric IWSAN. |
| $i$, $i = 1,2,\ldots M$ | Index for sensor and its corresponding actuator. |
| $p_i = p \in [0,1]$ | Probability for channel reliability. |
| T | Finite time horizon. |
| $t$, $t = 1,2,\ldots T$ | Index for TDMA slots. |
| $k_i = 1,2,3\ldots$ | Index number of $k^{th}$ sample packet from active sensor $i$. |
| $h_{t,i}(k_i)$ | Age of Information of the sample packet $k_i$ from sensor $i$ at the beginning of slot $t$. |
| $c_i(k_i - 1)$ | Action execution time based on $(k_i - 1)^{th}$ sample from sensor $i$. (Useful age). |
| $L_{t,i}(k_i)$ | Latency of $k_i^{th}$ sample at slot $t$ before its successful processing. |
| $P_i(k_i)$ | Processing time of the controller for the of $k_i^{th}$ sample from sensor $i$. (In this paper it aways takes the value 1). |
| $D_i(k_i)$ | Absolute Deadline for $k_i^{th}$ sample packet from sensor $i$. |
| $d_i(k_i)$ | Relative deadline for $k_i^{th}$ sample packet from sensor $i$. |
| $\vec{I}$ | Initial condition vector. |
| $\alpha_i = \alpha > 0$ $\forall i \in [1, M]$ | Real valued weight assigned to any sensor $i$ in the WSAN. |
| $AoI_{t,i}$ | Total Age incurred by the sensor $i$ during slot $t$. |
| $AoI_i$ | Total Age incurred by the sensor $i$ during the total finite time horizon T. |
| $\overline{AoI}$ | Average age of information in the network for M sensor-actuator pairs and finite time horizon T. |
| AL | Average latency of sample packets in the network for M sensor-actuator pairs and time horizon T. |
| $\overline{JT}$ | RMS jitter of the actuators in the network for M sensor-actuator pairs and finite time horizon T. |
| $U_{t,i}(k_i)$ | Utility of the information content in a sample $k_i$ from a sensor $i$ at any slot $t$. |
| $F_{t,i}(k_i)$ | Freshness of the information content in a sample $k_i$ from a sensor $i$ at any slot $t$. |
| $LX_{t,i}(k_i)$ | Laxity of sample $k_i$ from a sensor $i$ at any slot $t$. |
| $X_{t,i}(k_i)$ | Influence of laxity for any sample $k_i$ on its utilization at slot $t$.. |
| $Pri_{t,i}(k_i)$ | Priority of the active sensor $i$. |
| EXWSUoI | Expected weighted sum utility of information of the network. |
| $S'_t$ | Rearranged version of set $S_t$ arranged in decreasing order of priority of the constituent active sensors. |
| $R^{(t)}$ | Number of elements present in $S_t$ (or, $S'_t$). |
| $\{x_a\}_t$, $a = 1 \ldots R^{(t)}$ | Elements (active sensors) in $S_t$. |
| $\{e_{a'}\}_t$, $a' = 1 \ldots R^{(t)}$ | Elements in $S'_t$. |
| $Pri_{t,a}$ | Priority of the active sensor $x_a$ in $S_t$. |
| $Pri_{t,a'}$ | Priority of the active sensor $e_{a'}$ in $S'_t$. |
| $U_{t,a}$ | Utility of the active sensor $x_a$ in $S_t$. |
| $U_{t,a'}$ | Utility of the active sensor $e_{a'}$ in $S'_t$. |
| $N_t$ | Set of newly active sensors at the beginning of slot $t$. |
| $r^{(t)}$ | Number of elements present in $N_t$. |
| $\{y_n\}_t$, $n = 1 \ldots r^{(t)}$ | Elements in $N_t$. |
| $\omega^t$ | Active sensor sample scheduled for transmission at any slot $t$. |
| $drop^t$ | Critical sample dropped at the end of this slot $t$. |
| $V_t$ | Sum of utilization of all the active sensor samples present in $S_t$ (or, $S'_t$). |

**Devarpita Sinha** received her B. Tech. Degree in Electronics and Communication Engineering from West Bengal University of Technology, Kolkata, India, in 2014 and M.E. degree with First Class (Distinction) in the specialization if Wireless Communication from the Department of Electronics and Communication Engineering, Birla Institute of Technology (BIT) Mesra, Ranchi in 2016. She is currently pursuing her Ph.D.

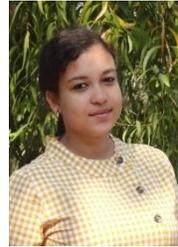

degree in the Department of Electronics and Electrical Communication Engineering, Indian Institute of Technology (IIT), Kharagpur, India. Her research interests include resource allocation problems, performance evaluation, and control in communication networks, scheduling and decision making, etc.

**Dr. Rajarshi Roy** received his B.E. degree with first-class (Hons.) in Electronics and Tele-Communication Engineering from Jadavpur University, Kolkata, West Bengal, India, in 1992, MSc. (Engineering) degree from the Department of Electrical Communication Engineering, IISc, Bangalore, India in 1995, and Ph.D. from Electrical and Computer Engineering Department, Polytechnic University,

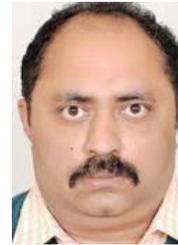

Brooklyn, New York, USA (Currently known as New York University, Tandon School of Engineering) in the year 2001. He is currently working as an Associate Professor in the Department of Electronics and Electrical Communication Engineering, Indian Institute of Technology, Kharagpur, India. His research interests include queuing theory, Markov decision theory, 5G/6G wireless communications, complex communication networks, massive MIMO, network coding, cooperative communication, cyber-physical systems, cognitive technology, etc.